\documentclass[showpacs,twocolumn,pre]{revtex4}

\usepackage[dvips]{epsfig}

\newcommand{\clL}{{\cal L}}

\newcommand{\tH}{\tilde{H}}
\newcommand{\tM}{\tilde{M}}

\newcommand{\prt}{\partial}

\newcommand{\vep}{\varepsilon}
\newcommand{\be}{\begin{equation}}
\newcommand{\ee}{\end{equation}}
\newcommand{\bea}{\begin{eqnarray}}
\newcommand{\eea}{\end{eqnarray}}

\begin{document}
\title{Application of hyperbolic scaling for calculation of
reaction-subdiffusion front propagation}

\author{A. Iomin${}^1$ and I.M. Sokolov${}^2$}

\affiliation{${}^1$ Department of Physics, Technion, Haifa 32000,
Israel \\
%
%
${}^2$ Institut f\"ur Physik, Humboldt-Universit\"at zu Berlin,
Newtonstrasse 15, 12489 Berlin, Germany}
\date{\today}

\begin{abstract}

A technique of hyperbolic scaling is applied to calculate a
reaction front velocity in an irreversible autocatalytic
conversion reaction $A+B\,\rightarrow\, 2A$ under subdiffusion.
The method, based on the geometric optics approach is a
technically elegant observation of the propagation front failure
obtained in Phys. Rev. E {\bf 78}, 011128 (2008).

\end{abstract}


\maketitle

\textit{Introduction}.- The problem of front propagation in
reaction-transport equations is attracting much attention that is
related to the considerable progress in our understanding of this
phenomenon via the generalization of the standard
reaction-diffusion scheme in the framework of the
Fisher-Kolmogorov-Petrovskii-Piskunov (FKPP) equation for
fractional reaction-subdiffusion systems
\cite{YH,Shkilev,Henry,fedotov2010a,sokolov,Abad,Campos2,Nec,sh,sokolov1,e031101}.
The description of reactions under subdiffusion is relevant to
strongly inhomogeneous environments, in porous media such as
certain geological formations or gels, in crowded cell interiors,
and so on. Another success in this field relates to developing an
appropriate new technique of treating front propagation, where an
appropriate hyperbolic scaling of the reaction-transport equation
makes it possible to estimate the overall rate of the spreading
reaction wave without resolving its shape \cite{Fr,F}. The method
of hyperbolic scaling is based on the introduction of a small
parameter $\vep\rightarrow 0$, and rescaling of coordinates and
time $(x,t)\rightarrow (x/\vep,t/\vep)$, and the contaminant's
density distribution function. In this case, the problem of the
wave propagation reduces to the dynamics of the leading edge, or
the reaction front. Therefore, one analyzes the reaction-transport
behavior in the leading edge, where, in the long-range and
long-time limits, the detailed shape of the travelling wave is not
important.

In this Report we demonstrate the hyperbolic scaling technique to
calculate the reaction front velocity in an irreversible
autocatalytic conversion reaction $A+B\,\rightarrow\, 2A$ under
subdiffusion, which, in the case of normal diffusion, is described
by the FKPP equation \cite{KPP,Fisher}. The present result is an
alternative Hamilton-Jacobi approach via hyperbolic scaling, which
is a more elegant presentation of the propagation front failure
observed in Ref.~\cite{sokolov}.

Recent results show that, contrary to normal diffusion, the
minimal propagation velocity is zero \cite{sokolov,Campos2,Nec},
which was interpreted as propagation failure (a general discussion
of this issue can be found in \cite{sh}). The main focus was on
situations when subdiffusion can be modelled within a
continuous-time random walk (CTRW) scheme with a waiting-time
probability density function (pdf) decaying according to the power
law, $\psi(t)\sim t^{\alpha}$, with $0<\alpha<1$. Analytical and
numerical calculations \cite{sokolov,sokolov1} corroborated this
picture, and in the regime of small reaction rates, for which the
continuous description applies, the front velocity was observed to
go as $t^{\frac{\alpha-1}{2}}$. Crossover arguments, presented in
\cite{e031101}, also support this picture.

\textit{Reaction-transport equation}.- The FKPP equation describes
a front propagating into the unstable state in bimolecular
autocatalytic conversion $A+B\,\rightarrow 2A$. Initially, the
whole system consists of particles  of type $B$. The introduction
of the A-individuals into some bounded spatial domain leads to the
propagation of a front of $A$ into the $B$-domain. A general
reaction-transport scheme that corresponds to the irreversible
$A+B\,\rightarrow\, 2A$ reaction process can be described by the
following generalization of the FKPP equation
\cite{YH,Shkilev,Henry,fedotov2010a,Abad,sokolov,VR,fir2011}
\bea\label{eq14} %
\frac{\prt B(x,t)}{\prt t}&=
\frac{a^2}{2}\int_0^t\Delta\left\{M(t-t')B(x,t')\right. \nonumber \\
&\times\exp\left.\left[-k\int_{t'}^t[1-B(x,t'{'})]dt'{'}\right]\right\}dt'
\nonumber \\
&-k[1-B(x,t)]B(x,t)\, .  %
\eea  %
Here $B(x,t)$ is a concentration of particles $B$ with the initial
condition $B(x,t=0)=B_0=1$, and the condition of the mass
conservation is
\[A(x,t)=1-B(x,t)\, .\]
The time kernel $M(\tau)$ is determined by the waiting time pdf in
the Laplace domain $\tilde{\psi}(u)=\hat{\clL}[\psi(t)]$
\be\label{rte_1}  %
\tilde{M}(u)=\frac{u\tilde{\psi}(u)}{1-\tilde{\psi}(u)}\,  . \ee %

\textit{Hyperbolic scaling}.- In  the sequel, we are concerned
with the front propagation of particles/individuals of the type
$A$. As mentioned above, to analyze the behavior of the leading
edge, we use the technique of hyperbolic scaling developed in
\cite{Fr}, see also \cite{F}, where the basic idea is that, in the
long-range and long-time limit, the detailed shape of the
travelling wave is not important, and the problem of wave
propagation corresponds to the dynamics of the leading edge or the
reaction front. We follow the details of the analysis presented in
Refs. \cite{fir2011,FI}. It is convenient to rewrite Eq.
(\ref{eq14}) using the variable change in the integration with the
memory kernel $ M(t-t')\rightarrow M(t')$. Thus Eq. (\ref{eq14})
for type $A$ individuals reads
\bea\label{front1_1a} %
\frac{\prt A}{\prt t}&=&
\sigma^2\int_0^t\Delta\Big\{M(t')(A(x,t-t')-1) \nonumber \\
&\times&\exp\Big[-k\int_{t-t'}^t[A(x,t'{'})]dt'{'}\Big]\Big\}dt'
\nonumber \\
&+& k[1-A(x,t)]A(x,t)\, . \eea %
After simple manipulation with the second derivative over space,
it reads
\bea\label{front1} %
\frac{\prt A}{\prt t}&=& \sigma^2\int_0^tM(t')\Big\{\Delta
A(x,t-t')-k\nabla A(x,t-t')
\nonumber \\
&\times&\int_0^{t'}\nabla A(x,t-t'{'})dt'{'} -k(A(x,t-t')-1)
\nonumber \\
&\times&\int_0^{t'}\Delta A(x,t-t'{'})dt'{'}+k^2(A(x,t-t')-1)
\nonumber \\
&\times&[\int_0^{t'}\nabla A(x,t-t'{'})dt'{'}]^2\Big\} \nonumber \\
&\times&\exp\Big[-k\int_0^{t'}[A(x,t-t'{'})]dt'{'}\Big]dt'
\nonumber
\\
&+&k[1-A(x,t)]A(x,t)
\, , %
\eea  %
where $\sigma^2=a^2/2$ and we use the $t'{'}\rightarrow t-t'{'}$.
The objective here is to find the rate of the front propagation
$v$ without resolving the shape of the travelling waves. We use
hyperbolic scaling for the coordinates $x$ and time $t$
\[x\rightarrow \frac{x}{\vep} \, ,~~~~~t\rightarrow \frac{t}{\vep}\, , \]
and the rescaled density
\[A^{\vep}\left( x,t\right) =A\left(\frac{x}{\vep}.\,\frac{t}{\vep} \right)\, .\]
We write the density $A^{\vep}\left( x,t\right) $ in the
exponential form
\be\label{wkb} %
A^{\vep }\left(x,t\right) =A_0\exp \left[ -\frac{
G^{\vep}\left(x,t\right) }{\vep }\right]\, , %
\ee  %
where the non-negative function $G^{\vep}\left(x,t\right) $
describes the asymptotics of the density function and plays a very
important role in the theory of front propagation.

At the next step, we rescale $x$ and $t$ variables in Eq.
(\ref{front1}) to obtain
\bea\label{front1a} %
\vep\frac{\prt A^{\vep}}{\prt t}&=&
\sigma^2\int_0^{t/\vep}M(t')\Big\{\Delta A^{\vep}(x,t-t')
\nonumber
\\
&-&k\nabla A^{\vep}(x,t-t')\int_0^{t'}\nabla A^{\vep}(x,t-t'{'})dt'{'} \nonumber \\
&-&k(A^{\vep}(x,t-t')-1)\int_0^{t'}\Delta
A^{\vep}(x,t-t'{'})dt'{'} \nonumber  \\
&+&k^2(A^{\vep}(x,t-t')-1)
[\int_0^{t'}\nabla A^{\vep}(x,t-t'{'})dt'{'}]^2\Big\} \nonumber \\
&\times&\exp\Big[-k\int_0^{t'}[A^{\vep}(x,t-t'{'})]dt'{'}\Big]dt'
\nonumber \\
&+&k[1-A^{\vep}(x,t)]A^{\vep}(x,t)\, .
\eea  %
We take into account that at finite times in the limit
$\vep\rightarrow 0$ the exponent in the Eq. (\ref{front1a}) tends
to unity since $A^{\vep}(x,t'{'})\rightarrow 0$ exponentially fast
due to Eq. (\ref{wkb}). Namely,
\[\lim_{\vep \to 0}
\exp\Big[-k\int_0^{t'}[A^{\vep}(x,t-t'{'})]dt'{'}\Big] =1\,
.\] %
Derivatives of $A^{\vep}(x,t)$ yield the following expressions
\[\prt_tA^{\vep}=-(A_0/\vep)(\prt_tG^{\vep})\exp(-G^{\vep}/\vep)\,
,\] %
\[\Delta
A^{\vep}=\Big[(A_0/\vep^2)(\prt_xG^{\vep})^2-(A_0/\vep)(\prt_x^2G^{\vep})\Big]
\exp(-G^{\vep}/\vep)\, .\] %
We also take into account that the terms of the order of
$(A^{\vep}(x,t))^2$ in braces tend to zero faster than $\frac{\prt
A^{\vep}}{\prt t}$ and disappear in the limit $\vep\rightarrow 0$.
Keeping in mind this limit and substituting these expressions in
Eq. (\ref{front1a}), one obtains for $G^{\vep}\equiv
G^{\vep}(x,t)$
\bea\label{front1b} %
&-&\frac{\prt G^{\vep}(x,t)}{\prt t}=
\sigma^2\int_0^{t/\vep}M(t')\Big\{\Big[\Big(\frac{\prt
G^{\vep}}{\prt x}\Big)^2-\vep\frac{\prt^2 G^{\vep}}{\prt x^2}\Big]
\nonumber
\\
&\times&\exp\Big[\frac{G^{\vep}(x,t)}{\vep}-\frac{G^{\vep}(x,t-\vep
t')}{\vep}\Big] 
+k\Big[\Big(\frac{\prt G^{\vep}}{\prt x}\Big)^2-\vep\frac{\prt^2
G^{\vep}}{\prt x^2}\Big] \nonumber  \\
&\times&\int_0^{t'}
\exp\Big[\frac{G^{\vep}(x,t)}{\vep}-\frac{G^{\vep}(x,t-t'{'})}{\vep}\Big]dt'{'}
\Big\}dt'  \nonumber \\
&+& k\Big[1-A^{\vep}(x,t)\Big]\, . %
\eea%
It follows from (\ref{wkb}) that as long as the function
\begin{equation}
G\left(x,t\right) =\lim_{\vep \to 0}G^{\vep }\left(x,t\right)
\end{equation}
is positive, the rescaled density $A^{\vep}\left(x,t\right)
\rightarrow 0$ as $\vep\rightarrow 0$. It also follows in this
limit that $\prt_x G(x,t-\vep t')=\prt_x G(x,t)$ and
\[\lim_{\vep\to 0}\left[\frac{G^{\vep}(x,t)}{\vep}-\frac{G^{\vep}(x,t-\vep
t')}{\vep}\right]=[\prt_tG(x,t)]t'\, .\]
Taking into account this limit expression, we obtain the following
equation for $G(x,t)$
\bea\label{front2} %
&\frac{\prt G}{\prt t}=-\sigma^2\Big(\frac{\prt G}{\prt
x}\Big)^2\int_0^{\infty}M(t') \exp\Big[\frac{\prt G}{\prt
t}t'\Big]dt' \nonumber \\
&-k\sigma^2\Big(\frac{\prt G}{\prt
x}\Big)^2\int_0^{\infty}M(t')\int_0^{t'}\exp\Big[\frac{\prt
G}{\prt t}t'{'}\Big]dt'{'}dt' -k\, . %
\eea %
In what follows $G(x,t)$ is the action, or Hamilton's principle
function, such that
\be\label{HJE_1a}  %
H=-\frac{\prt G}{\prt t}\, , ~~~~p=\frac{\prt G}{\prt x} \ee %
are the Hamiltonian, and the momentum, respectively. Therefore, it
follows from Eq. (\ref{HJE_1a}) that Eq. (\ref{front2}) is a kind
of Hamilton-Jacobi equation. The Laplace transform of the memory
kernel yields
\be\label{laplace} %
\tM(H)=\hat{\clL}M(t)\int_0^{\infty}M(t')e^{-Ht'}dt'=
\frac{H\tilde{\psi}(H)}{1-\tilde{\psi}(H)}\, , %
\ee %
where $\tilde{\psi}(H)=\hat{\clL}\psi(t)$ is the Laplace image of
the waiting time pdf [see Eq. (\ref{rte_1})]. Finally, one obtains
that the Hamiltonian can be found from the Hamilton-Jacobi
equation
\bea\label{HJE} %
\frac{\prt G}{\prt t}&=&-\sigma^2\Big(\frac{\prt G}{\prt
x}\Big)^2\tM(H) \nonumber \\
&+&\frac{k\sigma^2}{H}\Big(\frac{\prt G}{\prt
x}\Big)^2\Big[\tM(H)-\tM(0)\Big]-k\, . %
\eea %
Eventually, it reads
\be\label{ham} %
H=\sigma^2p^2\tM(H)-\frac{k\sigma^2}{H}p^2\Big[\tM(H)-\tM(0)\Big]+k\, ,  %
\ee%
and the action is
$G(x,t)=\int_0^t[p(s)\dot{x}(s)-H(p(s),x(s))]ds$.

The rate $v$ at which the front moves is determined from Eq.
(\ref{wkb}) at the condition $G(x,t)=0$. Together with the
Hamilton equations, this yields
\be\label{front3} %
v=\dot{x}=\frac{\prt H}{\prt p}\, ,~~~v=\frac{H}{p}\, . %
\ee %
Note that the first equation reflects the dispersion condition,
while the second one is a result of the asymptotically free
particle dynamics, when the action is $G(x,t)=px-Ht$. Taking into
account $x=vt$, one obtains Eq. (\ref{front3}) (see also details
of this discussion \textit{e.g.} in Refs. \cite{F,CFM}). Now we
analyze these two Eqs. (\ref{front3}) to define $v$.

\textit{Markovian case}.- \label{sec:markov}

First let us check the Markovian case, when
$\psi=\frac{1}{\tau}\exp(-t/\tau)$. Thus one has
\[\tilde{\psi}(H)=\frac{1}{1+H\tau}\, ,\]
where $\tau$ is a characteristic time scale. Therefore, from Eq.
(\ref{laplace}) the Laplace image of the memory kernel reads
\be\label{MarkovM} %
\tM(H)=\tM(0)=1/\tau\, . %
\ee %
In this case, the Hamiltonian in Eq. (\ref{ham}) is $H=Dp^2+k$,
where $\sigma^2/\tau=D$ is a diffusion coefficient. The moment is
\[p(H)=\sqrt{\frac{1}{D}(H-k)}\, .\]
From Eqs. (\ref{front3}) we have $2Dp=H/p$ and $H=2k$, which
yields for the overall velocity of the front propagation
\be\label{MarkovU} %
v=2\sqrt{kD}\, , \ee %
which is the classical FKPP propagation speed (see discussions in
Refs. \cite{sokolov,FI}).

\textit{Subdiffusion}.-
We have for subdiffusion $\psi=\frac{1}{1+(t/\tau)^{1+\alpha}}$,
which yields \cite{sokolov,FI}
\be\label{subdiff} %
\tM(H)=\frac{H^{1-\alpha}}{\tau^{\alpha}} \, ,~~(\tM(0)=0)\, , %
\ee %
where the transport exponent is defined in the range $0<\alpha<1$.
The Hamiltonian is
\be\label{ham1} %
H=D_{\alpha}(H-k)H^{-\alpha}p^2+k\, , %
\ee  %
where $D_{\alpha}=\sigma^2/\tau^{\alpha}$ is a generalized
diffusion coefficient From Eq. (\ref{ham1}) one obtains
\be\label{momentum} %
p(H)=\sqrt{\frac{H^{\alpha}}{D_{\alpha}}}\, , %
\ee  %
and Eqs. (\ref{front3}) result in
\be\label{clu1} %
\frac{\prt H}{\prt p}-
\frac{H}{p}=0=\sqrt{D_{\alpha}}(\frac{2}{\alpha}-1)H^{1-\frac{\alpha}{2}} \, . %
\ee %
This equation has the solution for $H=\tH=0$. Therefore, the
asymptotic velocity of the front propagation is
\[v(\tH)=0\, .\]
%

\textit{Conclusion}.- We demonstrated the technique of hyperbolic
scaling for the calculation of the reaction front velocity in an
irreversible autocatalytic conversion reaction $A+B\,\rightarrow\,
2A$ under subdiffusion. This is a technical presentation of the
powerful Hamilton-Jacobi method for asymptotic estimatimation of
the propagation front velocity observed in Ref.~\cite{sokolov}.

It should be admitted that the dispersion velocity $v(H$) also
determines the relaxation rate at the large time asymptotic $t
\rightarrow \infty$ for the \textit{finite} value of $H$. A
qualitative crossover argument based on the truncated power-law
distribution was suggested in Ref. \cite{e031101}. According to
the arguments, for short times the behavior of the velocity $v(H)$
must be similar to that in subdiffusion (it does not "feel" the
cutoff), whereas for long times the behavior is the classical one
with a constant minimal velocity, and there has to be a crossover
(no jump!) at a time $t_{\rm cr}$ between the two of them.
Assuming that the time dependence of the velocity in the anomalous
domain is $v(t)\sim t^{\beta}$ and, after the crossover to the
"normal" domain, the velocity, determined by Eq. (\ref{MarkovU}),
$v=2\sqrt{kD}$ is attained, both can be equated at $t_{\rm cr}$ to
obtain the corresponding $\beta$. To determine the crossover time,
it is plausible to argue with the amount of performed steps, as a
measure of mobility, which for the normal regime is $n_D(t)\propto
t/t^{1-\alpha}T^{\alpha}$ and in the subdiffusive regime reads
$n_{SD}(t)\propto (t/t_0)^{\alpha}$, equating them, one finds
$v(t<t_{\rm cr})\sim t^{\frac{1-\alpha}{2}}$.

The hyperbolic scaling also corroborates the relaxation picture
for the velocity $v(H)\sim t^{\frac{\alpha-1}{2}}$ obtained in
Refs. \cite{sokolov1,e031101}. An important point when considering
the relaxation in the framework of hyperbolic scaling is that the
process of relaxation for a subdiffusive front can be treated for
the finite energy $H=2k$ in the framework of the Markovian case.
For normal diffusion, the hyperbolic scaling method is rigourously
justified \cite{Fr}, and the method yields a correct result for
$v(H)$ in Eq. (\ref{MarkovU}), which is exactly the FKPP case.
This can be demonstrated for a truncated waiting time pdf
$\psi_T(t)$. The latter is convenient to take, \textit{e.g.}, in
the following power-law form
\be\label{psi2}  %
\psi_T(t)=\frac{\Big[1+(t_0/T)^{\alpha}\Big]e^{-t/T}}{1+(t/t_0)^{\alpha+1}}\,
,  \ee %
where $T$ has the role of the cutoff. Therefore, for any finite
$T$ the mean waiting time is finite:
\[\tau=\frac{\alpha t_0^{\alpha}}{1+(t_0/T)^{\alpha}}T^{1-\alpha}\,
.\]  %
For this normal diffusion hyperbolic scaling yields Eq.
(\ref{MarkovU}) for the velocity in the form
\be\label{vT} %
v(T)\propto\sqrt{Dk}\propto T^{\frac{\alpha-1}{2}}\,  ,  \ee   %
which corresponds to the relaxation rate obtained in
Ref.~\cite{e031101}.

Another specific property of the method is an effective
linearization of the generalized FKPP Eq. (\ref{front1}). It
should be admitted that this relates to considering a wavefront
with an exponentially decaying leading edge moving with a constant
velocity $v$. The standard, traditional way to perform this
analysis is first to linearize the equations. Hyperbolic scaling
performs it automatically, since the density $A$ is not zero only
when $G=0$. Moreover, it also affects the integrand kernel in Eq.
(\ref{front1}), namely, as admitted above, in the limit
$\vep\rightarrow 0$, the exponent in the Eq. (\ref{front1a}) tends
to unity since $A^{\vep}(x,t'{'})\rightarrow 0$ exponentially fast
due to Eq. (\ref{wkb}). This essential simplification makes it
possible to apply the strong machinery of the Laplace transform
and arrive at the analytically treatable Hamilton-Jacobi equation
(\ref{HJE}) that is an easy and elegant way to obtain the front
propagation, namely the failure of the latter. This nonlinear
kernel was also studied in relation to a mechanism coupling the
waiting time distributions to the reaction \cite{Campos2} to
resolve a controversy about reaction-subdiffusion front
propagation. To this end, a more general scheme of the local
waiting time was suggested \cite{Campos2} that eventually leads to
a more complicated analysis in the framework of the
Hamilton-Jacobi approach than presented here.

\end{document}